\documentclass [aps,pra,twocolumn,superscriptaddress]{revtex4}
\usepackage{amsmath}
\usepackage{amssymb}
\usepackage{graphicx}
\usepackage[T1]{fontenc}
\usepackage{color}

\usepackage{soul}

\begin{document}

\title{Mie-excitons: understanding strong coupling in dielectric nanoparticles}

\author{C. Tserkezis}
\email{ct@mci.sdu.dk}
\affiliation{Center for Nano Optics, University of Southern Denmark, Campusvej 55, DK-5230 Odense M, Denmark}
\author{P. A. D. Gon\c{c}alves}
\affiliation{Center for Nano Optics, University of Southern Denmark, Campusvej 55, DK-5230 Odense M, Denmark}
\affiliation{Center for Nanostructured Graphene, Technical University of Denmark, DK-2800 Kongens~Lyngby, Denmark}
\affiliation{Department of Photonics Engineering, Technical University of Denmark, DK-2800 Kongens~Lyngby, Denmark}
\author{C. Wolff}
\affiliation{Center for Nano Optics, University of Southern Denmark, Campusvej 55, DK-5230 Odense M, Denmark}
\author{F. Todisco}
\affiliation{Center for Nano Optics, University of Southern Denmark, Campusvej 55, DK-5230 Odense M, Denmark}
\author{K. Busch}
\affiliation{Max-Born-Institut, Max-Born-Stra{\ss}e 2A, D-12489 Berlin, Germany}
\affiliation{Humboldt-Universit\"{a}t zu Berlin, Institut f\"{u}r Physik, AG Theoretische Optik \& Photonik, Newtonstra{\ss}e 15, D-12489 Berlin, Germany}
\author{N. A. Mortensen}
\email{asger@mailaps.org}
\affiliation{Center for Nano Optics, University of Southern Denmark, Campusvej 55, DK-5230 Odense M, Denmark}
\affiliation{Center for Nanostructured Graphene, Technical University of Denmark, DK-2800 Kongens~Lyngby, Denmark}
\affiliation{Danish Institute for Advanced Study, University of Southern Denmark, Campusvej 55, DK-5230 Odense M, Denmark}

\begin{abstract}
We theoretically analyse the hybrid Mie-exciton optical modes arising from the strong coupling of excitons in organic dyes or transition-metal dichalcogenides with the Mie resonances of high-index dielectric nanoparticles. Detailed analytic calculations show that silicon--exciton core--shell nanoparticles are characterised by a richness of optical modes which can be tuned through nanoparticle dimensions to produce large anticrossings in the visible or near infrared, comparable to those obtained in plexcitonics. The complex magnetic-excitonic nature of these modes is understood through spectral decomposition into Mie-coefficient contributions, complemented by electric and magnetic near-field profiles. In the frequency range of interest, absorptive losses in silicon are sufficiently low to allow observation of several periods of Rabi oscillations in strongly coupled emitter-particle architectures, as confirmed here by discontinuous Galerkin time-domain calculations for the electromagnetic field beat patterns. These results suggest that Mie resonances in high-index dielectrics are promising alternatives for plasmons in strong-coupling applications in nanophotonics, while the coupling of magnetic and electric modes opens intriguing possibilities for external control.
\end{abstract}
\maketitle


\section{Introduction}\label{Sec:intro}

The interaction of optical modes in a structured electromagnetic (EM) environment with the photons emitted by atoms, molecules, organic dyes, quantum dots or nanomaterial defects has long been in the forefront of interest in photonics, as it is characterised by novel fundamental physics and exciting applications \cite{drexhage_jlum1,kaluzny_prl51,rempe_prl58,thompson_prl68,reithmaier_nat432,wolters_apl97,albrecht_njp15}. Plasmons, in particular, are frequently combined with classical and quantum emitters, and are acknowledged as excellent templates for sensing \cite{anker_natmat7,galush_nl9}, fluorescence \cite{chance_acp37,fu_lpr3} and Raman enhancement \cite{xu_pre62}, and optical communications \cite{akimov_nat450,martin-cano_nl10}. Recently, strong coupling of emitters with surface plasmon polaritons in metal films or localised surface plasmons in nanoparticles (NPs) has turned into a rapidly growing field, due to its potential for applications in quantum optics \cite{torma_rpp78,necada_ijmpb31,gerhardt_ijmpb31,baranov_acsphoton5}. In so-called plexcitonic architectures, plasmons confine light to small volumes that largely overcome the diffraction limit \cite{benz_sci354}, dramatically enhancing the coupling strength and enabling light-matter interactions to enter the strong coupling regime, which is characterised by Rabi oscillations in the emitter occupation and hybrid optical states of mixed light-matter nature \cite{bellessa_prl93,dintinger_prb71,sugawara_prl97,fofang_nl8,gonzalez_prl110,chatzidakis_jmo65}. Nevertheless, full implementation of plasmonic designs in applications is still hindered by their high ohmic losses \cite{khurgin_natnano10,kewes_prl118}, and different schemes are explored.

Among the various materials encountered in nanophotonics, high-refractive-index dielectrics hold a prominent position, as they combine low loss with -- unavailable in metals -- magnetic Mie modes and compatibility with existing nanoelectronic platforms \cite{garcia_oex19,evlyukhin_nl12,staude_natphot11}. Silicon NPs, in particular, have been successfully exploited as building blocks for nanoantennas, sensing environments and optical metamaterials \cite{staude_acsnano7,baranov_opt4}. These advances have shown that on many occasions dielectric NPs do not fall short in comparison to their plasmonic counterparts \cite{kuznetsov_sci354,jahani_natnano11}. It is therefore natural to consider them as prospective templates for strong coupling designs. Recently, a numerical study showed that resonance coupling in dye--silicon heteroaggregates is indeed possible \cite{wang_nl16}, but a thorough investigation and theoretical understanding of the nature of the resulting hybrid polaritons is still missing.

Here we present a theoretical study of the hybrid Mie-exciton modes that arise from the interaction of magnetic dipolar modes in silicon nanospheres and nanoshells with the excitons sustained by $J$-aggregates of organic molecules or two-dimensional (2D) transition-metal dichalcogenides (TMDs) \cite{goncalves_prb97,cuadra_nl18}. Through analytic solutions based on Mie theory \cite{Bohren}, combined with discontinuous Galerkin time-domain (DGTD) dynamical studies \cite{konig_pnfa8}, we show that mode splittings of the order of 150--200\,meV can be achieved by coupling emitters to dielectric NPs. Far- and near-field analysis shows that the resulting modes are characterised by a mixed, electric-magnetic dipole nature, while the periods of the EM field temporal oscillations (analogous to Rabi oscillations in two-level systems) are in perfect agreement with the frequency splitting. Comparison with plasmonic architectures with similar spectral anticrossings shows that in the latter the beat patterns decay much faster. Mie-excitons are therefore efficient alternatives to plexcitonics, thereby envisaging low-loss and externally controllable strong-coupling templates for applications in nanophotonics.

\section{Results and discussion}\label{Sec:results}

\begin{figure}[h]
\centerline{\includegraphics*[width=1\columnwidth]{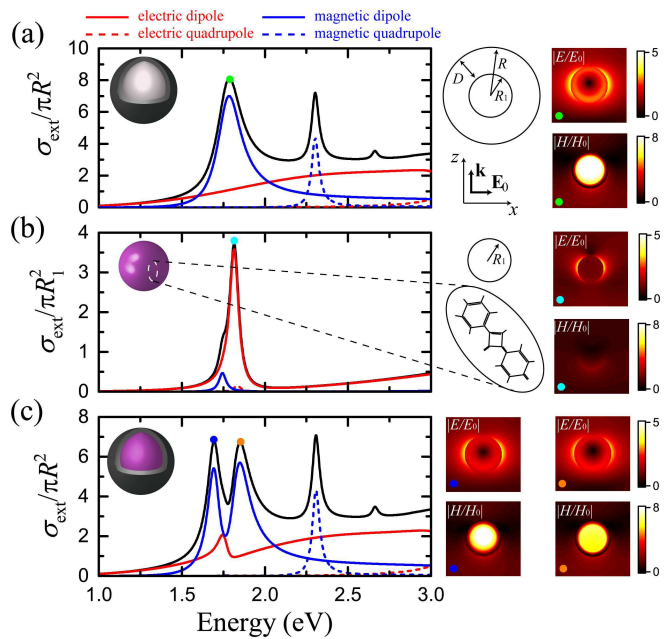}}
\caption{(a) Normalised extinction cross section ($\sigma_{\mathrm{ext}}$, black line) of the uncoupled [$f = 0$ in Eq.~(\ref{Eq:Lorentz}) for the core, $\varepsilon = \varepsilon_{\mathrm{Si}}$ for the shell] core--shell NP of the inset ($R_{1} = 70$\,nm, $R =$ 100\,nm), and the contributions from the electric (red) and magnetic (blue) dipolar (solid lines) and quadrupolar (dashed lines) Mie coefficients. Electric and magnetic field amplitude profiles (in the $xz$ plane, normalised to the incident fields), $|E/E_{0}|$ and $|H/H_{0}|$ respectively, at the frequency of the magnetic dipolar resonance (green dot in the spectra) are shown on the right-hand side, for an incident plane wave propagating along the $z$ and polarised along the $x$ axis. (b) Same as in (a), for an excitonic sphere ($R_{1} = 70$\,nm) described by the dielectric function of Eq.~(\ref{Eq:Lorentz}). (c) Analysis of $\sigma_{\mathrm{ext}}$, and corresponding field profiles, for the coupled [$f = 0.4$ in Eq.~(\ref{Eq:Lorentz})] exciton core--silicon shell NP ($R_{1} = 70$\,nm, $R =$ 100\,nm).
}\label{fig1}
\end{figure}

The silicon--exciton NPs considered here comprise a homogeneous spherical core of radius $R_{1}$ and a concentric shell of thickness $D$, so that the total radius is $R = R_{1} + D$, as schematically shown in Fig.~\ref{fig1}(a). Both arrangements of core and shell materials are allowed, while air is the host medium throughout the paper. Silicon permittivity ($\varepsilon_{\mathrm{Si}}$) follows the experimental values of Ref.~\cite{green_semsc92}, so as to provide a dispersive and lossy description of its optical response (silicon loss is negligible in the infrared, but cannot be ignored in the visible). For the generic excitonic material we use a Lorentz dielectric function,
\begin{equation}\label{Eq:Lorentz}
\varepsilon_{\mathrm{exc}} = \varepsilon_{\infty} - \frac{f \omega_{\mathrm{exc}}^{2}}{\omega^{2} - \omega_{\mathrm{exc}}^{2} + \mathrm{i} \omega \gamma_{\mathrm{exc}}}~,
\end{equation}
where $\omega_{\mathrm{exc}}$ is the excitonic transition frequency, $\gamma_{\mathrm{exc}}$ the corresponding damping rate, $f$ the oscillator strength, and $\varepsilon_{\infty}$ the background permittivity. Choosing $\hbar \omega_{\mathrm{exc}} = 1.76$\,eV, $\hbar \gamma_{\mathrm{exc}} = 0.05$\,eV, $f = 0.4$, and $\varepsilon_{\infty} = 3$ accounts fairly well for the resonance in the dielectric function of squaraine \cite{cacciola_acsnano8} while disregarding broadening at higher energies to facilitate theoretical analysis.

We first consider an excitonic core--silicon shell NP with $R_{1} = 70$\,nm and $R = 100$\,nm. Such a design, though challenging from an experimental point of view, is beneficial to theoretical understanding, as it reduces mode mixing. Indeed, as can be seen by decomposing the extinction cross section ($\sigma_{\mathrm{ext}}$, normalised to the geometrical cross section) in Fig.~\ref{fig1}(a), when the excitonic resonance of the core is disregarded [$f = 0$ in Eq.~(\ref{Eq:Lorentz})], the silicon shell is characterised by well-defined magnetic Mie resonances of increasing multipolar order (blue solid and dashed lines for the dipolar and quadrupolar modes, respectively), over the tail of a wide but weak electric dipolar background (red solid line). This is in contrast to the case of homogeneous silicon spheres, where the electric dipole mode is pronounced and strongly overlaps with the low-energy magnetic dipole one \cite{garcia_oex19}. The magnetic dipole nature of the first mode, at about 1.79\,eV, is further confirmed by the electric and magnetic field amplitude profiles ($|E/E_{0}|$ and $|H/H_{0}|$, normalised to the corresponding incident field) on the right-hand side.

This almost negligible contribution of the electric dipole mode allows to verify whether the magnetic dipole mode of the silicon shell can significantly interact with the electric dipole of the excitonic core, once tuned to coincide in frequency. The extinction spectrum of the bare excitonic core ($R_{1} = 70$\,nm), described by the dielectric function of Eq.~(\ref{Eq:Lorentz}), is shown in Fig.~\ref{fig1}(b). For such large size, a second, magnetic dipolar mode is also excited, and appears as a shoulder in the extinction spectrum. A smaller radius ensures excitation of a single electric dipole mode (see also near-field profiles on the right-hand side), but our calculations showed that in that case ($R_{1} \lessapprox 40$\,nm) the core dipole moment is not strong enough to support the desired coupling strength.

When the two components are merged in a core--shell geometry, as depicted in Fig.~\ref{fig1}(c), their modes couple like harmonic oscillators \cite{torma_rpp78}, leading to a double-peak spectrum where the hybrid Mie-exciton modes are separated by a split of 156\,meV, comparable to the linewidth of the silicon shell resonance in Fig.~\ref{fig1}(a) (145\,meV). These modes combine the electric and magnetic characters of the uncoupled components, but maintain a dominantly magnetic dipolar nature due to the relative differences in the dipole strengths of the original modes. This is verified both by the Mie-coefficient analysis and the near-field profiles on the right-hand side. The field profiles are not identical, because of the asymmetric electric dipole background provided by the silicon shell [Fig.~\ref{fig1}(a)], but the magnetic field enhancement inside the otherwise non-magnetic NP core is evident. A more symmetric extinction spectrum with equal linewidths for the two hybrid modes can be achieved by fine, sub-nm tuning of radii.

\begin{figure}[h]
\centerline{\includegraphics*[width=1\columnwidth]{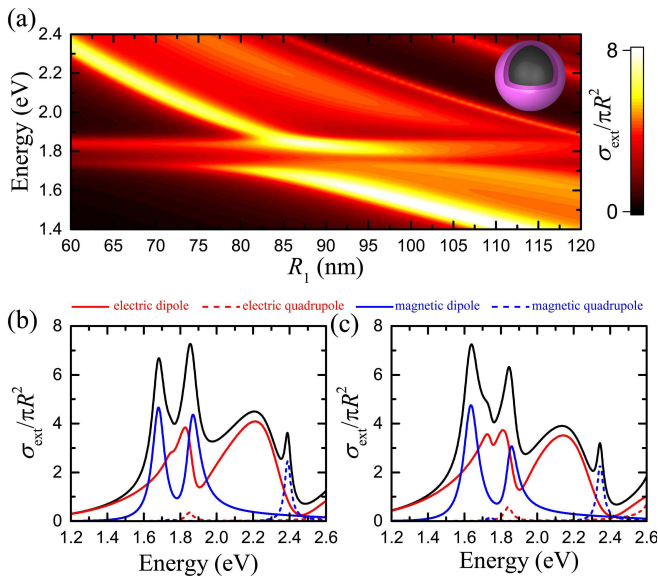}}
\caption{(a) Extinction colour map as a function of the core radius $R_{1}$ of the silicon core--exciton shell nanosphere shown in the inset ($D = 20$\,nm). Anticrossing is obtained for $R_{1} \simeq 85$\,nm. (b) Mie-coefficient analysis of the anticrossing spectra ($R_{1} = 85$\,nm, $R = 105$\,nm). (c) Same as (b), for a thicker excitonic shell ($R_{1} = 87$\,nm, $D = 30$\,nm).}\label{fig2}
\end{figure}

In what follows we invert the material arrangement and consider dielectric NPs covered with homogeneous excitonic shells. For thick shells, the NP sustains not only an exciton resonance at fixed frequency, but also modes originating from its negative dielectric function, that become stronger as the shell increases \cite{antosiewicz_acsphot1,gentile_jopt19}. Nevertheless, the optical response can still be precisely tuned by adjusting the silicon core. In Fig.~\ref{fig2}(a) we increase $R_{1}$ (for constant $D = 20$\,nm) to shift the magnetic dipole mode of the core from the infrared (large NPs) all the way to the visible. When it matches the excitonic resonance of the shell (for $R_{1} = 85$\,nm, with a linewidth of 97\,meV), an anticrossing of 178\,meV emerges [Fig.~\ref{fig2}(b)], indicating strong coupling. Wider splits, exceeding 200\,meV can be achieved by increasing the dipole moment of the excitonic layer, e.g., by increasing $f$ in Eq.~(\ref{Eq:Lorentz}), or, for fixed permittivity, by increasing the shell thickness, as shown in Fig.~\ref{fig2}(c) ($R_{1} = 87$\,nm, $D = 30$\,nm). However, in such situations the geometrical shell modes are more pronounced, appearing as third peaks or shoulders in the spectra, and the mode splitting is not well-defined \cite{antosiewicz_acsphot1}. Furthermore, in Fig.~\ref{fig2}(c) two different couplings can be identified in the Mie coefficients, with different splits in the magnetic and electric dipole contribution. This complex interaction originates from the fact that the shell now sustains two modes, close in frequency but with similar strengths and linewidths (spectra not shown here), which both interact with the silicon core. We also note that the middle peak is almost entirely absorptive, stressing the need to be careful when discussing strong coupling in such situations, as different conclusions might be drawn by examining scattering or absorption spectra \cite{zengin_jpcc120}.

\begin{figure}[h]
\centerline{\includegraphics*[width=1\columnwidth]{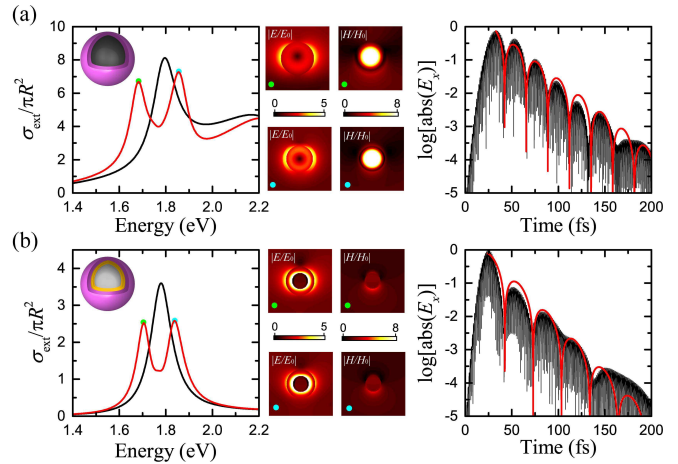}}
\caption{(a) Left panel: extinction spectra for the uncoupled [black line, $f = 0$ in Eq.~(\ref{Eq:Lorentz})] and coupled (red line) silicon--exciton core--shell NP of the inset. Middle panel: electric and magnetic field profiles in the $xz$ plane, for an $x$-polarised plane wave at the frequencies of the two coupled Mie-excitons (green and light-blue dots in the spectra). Right panel: time dependence of the logarithm of the $x$ component of the total (incident + scattered) electric field (black line), at 25\,nm from the NP surface along the $x$ axis, when the NP is irradiated by 
a Gaussian envelope with 16fs FWHM and plane-wave transversal profile. The red line represents the corresponding envelope function. NP dimensions are as in Fig.~\ref{fig2}(b). (b) Same as in (a), for a silica--gold--dye trilayered NP (silica core radius 19.5\,nm, $R_{1} = 25$\,nm, $D = 20$\,nm.}\label{fig3}
\end{figure}

A key element in strong coupling studies is the period and decay of Rabi oscillations in the occupation of two-level systems. For our excitonic layers such an occupation is not strictly defined, but important information can be retrieved by the time evolution of the EM field around the NP. The beat pattern for the $x$ component of the electric field ($E_{x}$) at 25\,nm from the NP surface, obtained with the DGTD method \cite{niegemann_pnfa7}, is shown by a black line on the right-hand side of Fig.~\ref{fig3}(a). For the dynamic calculations, the NP and surrounding air and perfectly matched layer termination were discretised into 19500 tetrahedral elements with third order Lagrange basis functions (element size $15\,$nm inside the NP, $50\,$nm in air). The setup was excited by a pulse with plane-wave transversal profile and a narrowband Gaussian envelope [carrier wavelength 700\,nm, full-width at half-maximum (FWHM) 16\,fs]. The obtained oscillations are accurately described by an envelope function (red line) of an exponentially decaying cosine whose frequency exactly matches the mode splitting ($\hbar \omega = 0.179$\,eV) of the left-hand spectra. Similar beat patterns were obtained for the magnetic field inside the NP. Interestingly, while the magnetic field profiles of the hybrid modes are similar, their electric field distributions are very different [middle panel in Fig.~\ref{fig3}(a)], most notably regarding the field enhancement inside the shell for the high-energy branch. In view of our discussion of Fig.~\ref{fig1}, the differences can be attributed to the asymmetric background originating from additional, higher-order modes in the frequency region of the split.

To explore similarities and advantages of silicon NPs over their plasmonic counterparts, we compare in Fig.~\ref{fig3} the extinction spectra, electric field profiles, and beat patterns, with those of an exciton-covered gold NP. As before, the dynamic response was simulated using DGTD with a similar mesh with 27500 elements (element size in the metal $6\,$nm) and modelling the dielectric properties of gold~\cite{johnson_prb6} through a Drude-Lorentz fit with maximal relative error of 6\% over the spectral range 400-2000\,nm (see \cite{Au_data} for numeric values; the same model was used in the frequency-domain calculations for consistency):
\begin{equation}\label{Eq:AuLorentz}
\varepsilon_{\mathrm{Au}} =
\varepsilon_{\infty}^{\mathrm{Au}} - \frac{\omega_{\mathrm{p}}^{2}}{\omega^{2} + \mathrm{i} \omega \gamma_{\mathrm{p}}} - 
\sum_{i=1}^{2} \frac{f_{i} \omega_{\mathrm{L}i}^{2}}{\omega^{2} - \omega_{\mathrm{L}i}^{2} + \mathrm{i} \omega \gamma_{\mathrm{L}i}} ~.
\end{equation}
For a good comparison, one should have the same geometrical parameters, extinction spectra (in terms of resonance position and linewidth), scattering and absorption contributions, and mode anticrossing. Since this is practically impossible, we focus here on a NP with similar linewidth and resonance position of its lower-energy mode before coupling -- in plasmonics, this is of course ian electric dipolar mode. To tune the plasmon mode at 1.76\,eV, we consider silica--gold nanoshells ($\varepsilon_{\mathrm{silica}} = 2.13$) instead of homogeneous gold spheres, and introduce plasmon hybridisation as a mode shifting mechanism \cite{tserkezis_nscale8,tserkezis_acsphoton5}. For a silica core of radius 19.5\,nm, total nanoshell radius ($R_{1}$ in the previous context) 25\,nm, and $D = 20$\,nm, the extinction spectrum of Fig.~\ref{fig3}(b) is obtained. When the exciton resonance is introduced [$f \neq 0$ in Eq.~(\ref{Eq:Lorentz})], a 136\,meV split is readily obtained. This frequency provides an envelope function that again excellently describes the temporal profile of the electric field (right-hand panel). More importantly, fewer [as compared to Fig.~\ref{fig3}(a)] periods of oscillations are observable before complete dissipation (a reduction of the field by 3 orders of magnitude is produced already after 100\,fs), which implies that dielectric NPs are more suitable candidates for dynamical studies of strong coupling. Another striking difference can be seen in the field profiles [middle panel in Fig.~\ref{fig3}(b)] in the plasmonic case, where a single electric dipole mode interacts with the exciton of the dye and the resulting hybrid modes are practically identical.

\begin{figure}[h]
\centerline{\includegraphics*[width=1\columnwidth]{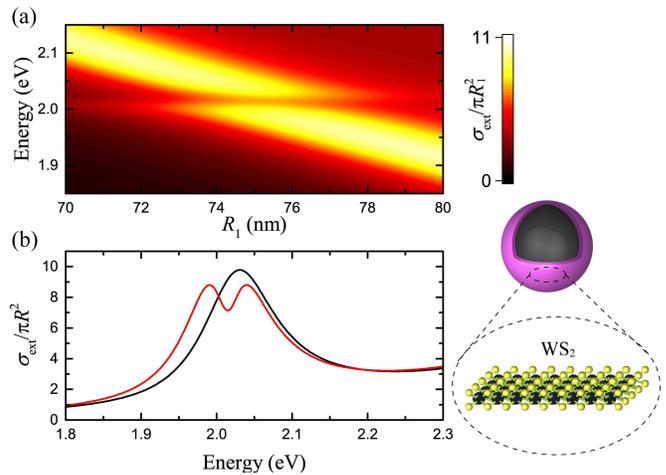}}
\caption{(a) Extinction contour as a function of the radius $R_{1}$ of the silicon core, for the silicon--WS$_{2}$-monolayer NP shown schematically on the right-hand side. (b) Bare silicon NP (black line) and WS$_2$-coated silicon NP (red line) extinction spectra near the crossing point ($R_{1} = 75$\,nm).}\label{fig4}
\end{figure}

Finally, as an alternative, more robust and easily achievable with current nanofabrication techniques design, we replace the shell of organic molecules with a monolayer of TMD, specifically WS$_{2}$, as was recently done in Ref.~\cite{lepeshov_acsami}. Atomically-thin semiconductors based on TMDs -- with chemical formula MX$_2$, where M=\{W,Mo\} and X = \{S,Se,Te\} -- exhibit strong light-matter interactions \cite{liu_natphot9,low_natmat16}, at the heart of which lie excitons with large binding energies due to the weaker screening resulting from the reduced dimensionality \cite{wang_rmp90}. The optical response of TMDs in the visible and near-infrared is dominated by excitonic resonances, even at room temperature \cite{li_prb90}, which, combined with their tuneability (provided by layer number, strain, gating, etc.), makes them attractive platforms for strong coupling under ambient conditions \cite{flatten_scirep6}. Fabrication of gold nanospheres coated with few-to-a-single MoS$_2$ layers has been experimentally demonstrated in a number of recent studies~\cite{li_nl16,lavie_nanotech28,distefano_nn11,chen_nscale10}. Furthermore, a monolayer of WS$_{2}$ deposited on a metal film was recently shown to lead to mode splittings of the order of 70--80\,meV, while even higher values can be achieved by engineering the dielectric environment \cite{goncalves_prb97}. For our purposes, such an excitonic material provides a "cleaner" system, with just a single excitonic resonance in the spectral window of interest. For the dielectric function of WS$_{2}$ we use the experimental data of \cite{li_prb90}, fitted with multi-oscillator Lorentzian, as described in \cite{goncalves_prb97}. Within Mie theory, the 2D material is introduced as a surface current and corresponding conductivity to the boundary conditions. In Fig.~\ref{fig4}(a) we show extinction spectra for WS$_2$-covered silicon NPs as the core radius is modified. Clearly, an avoided crossing emerges around the WS$_2$ excitonic resonance (i.e., at 2.01\,eV for the so-called $A$-exciton). The corresponding uncoupled and coupled spectra at the crossing point (for $R_{1} = 75$\,nm) are shown in Fig.~\ref{fig4}(b) (black and red line for the bare WS$_2$-covered silicon NP, respectively). Interestingly, the energy split is relatively narrow, of only about 50\,meV, as compared to the 100\,meV split of Ref.~\cite{lepeshov_acsami}, probably due to the difference in modelling the TMD layer. Size and shape engineering should enable stronger mode hybridisation, while 2D TMDs can even be used as substrates or covering layers of photonic nanostructures. Other possibilities include strain-engineering \cite{wu_nl18} and using few-layer TMDs to increase the effective interaction volume, although in the latter case this has to be judiciously balanced with the fact that the TMD bandgap becomes indirect as the layer number increases.

\section{Conclusions}\label{Sec:conclusion}

We explored the coupling of excitons in organic molecules and 2D TMDs with the magnetic Mie modes of silicon NPs. Such complex nanostructures are characterised by rich optical spectra, dominated by hybrid Mie-exciton modes, with splittings comparable to those seen in plexcitonics. Nevertheless, attention is required when analysing far-field spectra, as additional modes can mix with the Mie-excitons and manifest themselves in scattering, in absorption, or in both. Near-field profiles verify that these modes combine the electric and magnetic field enhancements of their constituents, thus promising easier external control with an applied bias or magnetic field. Direct comparison between the time evolution of the fields around silicon-- or gold--exciton core--shell NPs shows that dielectrics are more efficient substitutes for metals when studying strong coupling dynamics. Finally, we suggest alternative architectures based on TMD-covered high-index NPs, which provide more flexibility and tuneability in practical realisations.

\begin{acknowledgements}
We thank V.~A.~Zenin for stimulating discussions. N.~A.~Mortensen is a VILLUM Investigator supported by VILLUM FONDEN (grant No. 16498). The Center for Nano Optics is financially supported by the University of Southern Denmark (SDU 2020 funding). The Center for Nanostructured Graphene is sponsored by the Danish National Research Foundation (Project No. DNRF103). F.~Todisco and C.~Wolff acknowledge funding from MULTIPLY  fellowships under the Marie Sk\l{}odowska-Curie COFUND Action (grant agreement No. 713694). K.~Busch acknowledges support by the Deutsche Forschungsgemeinschaft within the frame of project B10 of SFB 951 (HIOS).
\end{acknowledgements}

\end{document}